\documentstyle [preprint,aps,epsf,floats]{revtex}

\global\arraycolsep=2pt            

\makeatletter
\def\footnotesize{\@setsize\footnotesize{9.5pt}\xpt\@xpt
\abovedisplayskip 10pt plus2pt minus 5pt
\belowdisplayskip \abovedisplayskip
\abovedisplayshortskip \z@ plus 3pt
\belowdisplayshortskip 6pt plus 2pt minus 2pt
\def\@listi{\topsep 6pt plus 2pt minus 2pt
\parsep 3pt plus 2pt minus 1pt \itemsep \parsep}}
\makeatother
\tightenlines

\begin {document}

\preprint {UW/PT-99-1}

\title {New Use of Dimensional Continuation
\\ 
Illustrated by $dE/dx$ in a Plasma
\\
and the Lamb Shift}

\author {Lowell S. Brown}

\address {Department of Physics, University of Washington
        \\Seattle, Washington 98195}

\date{\today}

\maketitle

\begin {abstract}
    {%
        Physical processes ranging from the Lamb shift to the energy
        loss $dE/dx$ of a charged particle traversing a plasma entail
	processes that occur over a wide range of energy or length 
	scales. Different physical mechanisms dominate at one or the
	other end of this range.  For example, in the energy loss
        problem,  soft collisions that are screened by collective 
	effects are important at large distances, while  
	at short distances hard collisions are important where the 
	exact details of the single-particle interactions 
	must be taken into account.  We introduce a novel application 
        of dimensional continuation. The soft processes dominate at
	all scales when the spatial dimension $\nu$ is less than
	$3$, and we use them to compute the result to leading order 
        for $\nu < 3$. On the other hand, the hard processes dominate 
	at all scales for $\nu > 3$, and we use them to compute the
	result to leading order for these spatial dimensions. We then
	explain why the sum of the analytic continuation of these 
	disparate mechanisms yields the correct leading-order result 
	for the physical limit at $\nu = 3$
	dimensions. After applying this new method to the energy loss 
	problem in some detail, we then show how it also provides a 
	very short and easy way to compute the Lamb shift.

    }%
\end {abstract}

\newpage

\section{Introduction}

The purpose of this paper is to introduce a new application of
dimensional continuation to physical problems that involve
simultaneously both short-distance, ultraviolet processes and
long-distance, infrared processes.  Such problems typically involve a
logarithm of a large ratio of two different scales.  Although the
overall coefficient of the logarithm is usually easy to compute, the
constant under the logarithm is often difficult to find.  Our new
method makes the computation of this constant under the logarithm
easy. We shall illustrate the simplicity of the method and its wide
range of applicability by using it to solve two very diverse physical
problems, the $dE/dx$ energy loss of a fast particle traversing a
fully ionized, but non-relativistic dilute plasma and the Lamb shift
of hydrogen-like atoms. We shall first work out the energy loss
problem in some detail because it involves rather elementary physics,
and a self-contained exposition can be presented within a short
space. The plasma example also does share some common features with
relativistic plasmas such as those that appear in QCD and weak
interaction physics.  Indeed, this simple problem serves as a useful
test of the validity of methods used in those more complex
problems. In this regard, it should be noted that although our
non-relativistic, Abelian example is conceptionally simpler than those
in the relativistic, non-Abelian gauge theory, the non-relativistic
plasma involves a Debye length and a plasma frequency that are quite
distinct parameters while, in the extreme relativistic theory, these
parameters become essentially the same (in lowest-order perturbation
theory.) Thus a verification of the treatment of different physical 
processes may be more difficult in the relativistic theory, and the
non-relativistic case may well serve as a quite useful testing bed
for methods used in the relativistic case. 
Here our aim is to explain our
method, and although we shall re-derive results that have been
obtained before, the derivations will clearly describe and illustrate
the power of the method. It has been applied recently to plasma
energy-loss problems in which the constants under the logarithms were
not known \cite{sawyer}, \cite{brown2}.

Dimensional regularization is widely employed in relativistic quantum
field theory to make otherwise divergent expressions finite and well
behaved, and then to implement the renormalization procedure.  It
should be emphasized that here we are making use of a different
and novel application of continuation to spatial dimensions $\nu \ne 3$ to
compute results that are always well-defined and finite at the
physical $\nu = 3$ dimension.\footnote{In the Lamb shift example at
the end of this paper, we work with renormalized quantities that are
well-defined and finite.} We are not using dimensional
continuation to render infinities finite so as to perform
renormalizations as one does in quantum field theory.  Moreover,
our purpose here is to introduce and describe this new application of
dimensional continuation by two very different but well-known 
physical problems in order to illustrate its range of application. 
What is new is the method. 

It is worth first illustrating our method with a trivial mathematical 
example, the behavior of the modified Hankel function $K_\nu(z)$ in 
the small argument $z$ limit with the index $\nu$ also small. The 
argument $z$ will play the
role of the small parameter in our work below; the index $\nu$ will
play the role of the dimensionality except that in this simple Bessel
function example we shall examine the region where $\nu$ is near zero,
not three.  In general, the Hankel function has the integral representation
\begin{equation}
K_\nu(z) = { 1 \over 2} \, \int_0^\infty {dk \over k} \, k^\nu
\exp\left\{ - {z \over 2} \, \left( k + {1 \over k} \right) \right\}
\,.
\label{int}
\end{equation}
Although $k$ is simply a dummy integration variable, it is convenient
to think of it as a wave number or momentum variable.  When $z$ is
small, $ \exp\left\{ - {z \over 2} \, \left( k + { 1 \over k} \right)
\right\} $ may be replaced by $1$ except when one or the other of the
factors $\exp\{ - z \, k /2 \}$ or $\exp\{ - z /(2 \, k) \}$ is needed
to make the $k$ integration converge in the neighborhood of one of its
end points. When $\nu$ is slightly less than zero, the integral
(\ref{int}) is dominated by the small $k$, ``infrared or
long-distance'', region.  In this case, only the $\exp\{ - z /(2 \, k)
\}$ factor is needed to provide convergence, and we have
\begin{eqnarray}
\nu < 0 \, : \qquad\qquad\qquad &&
\nonumber\\
K_\nu(z) &\simeq& { 1 \over 2} \, \int_0^\infty {dk \over k} \, k^\nu
\exp\left\{ - {z \over 2 \,k } \right\}
\,.
\end{eqnarray}
The variable change $ k = z / ( 2 t) $ places this integral in the form of
the standard representation of the gamma function, and we thus find
that the leading term for small $z$ in the region $ \nu < 0 $ is given
by
\begin{eqnarray}
\nu < 0 \, : \qquad\qquad\qquad &&
\nonumber\\
K_\nu(z) &\simeq& { 1 \over 2} \, \left( { z \over 2} \right)^\nu
\Gamma(- \nu) 
\nonumber\\ 
&\simeq&  -{ 1 \over 2 \nu } \, \left( {z \over 2} \right)^{\nu} 
( 1 + \nu \gamma )  \,,
\label{small}
\end{eqnarray}
where $\gamma = 0.5772 \cdots$ is Euler's constant.  Note that the
second line describes the behavior for $\nu < 0$ near $\nu = 0$ 
including the correct
finite constant as well as the singular pole term. 

When $\nu$ is
slightly greater than zero, the integral (\ref{int}) is dominated by
the large $k$, ``ultraviolet or short-distance'' regions. In this
case, only the $\exp\{ - z \, k /2 \}$ factor is needed to provide
convergence, and we have
\begin{eqnarray}
\nu > 0 \, : \qquad\qquad\qquad &&
\nonumber\\
K_\nu(z) &\simeq& { 1 \over 2} \, \int_0^\infty {dk \over k} \, k^\nu
\exp\left\{ - {z \, k \over 2}  \right\} \,.
\end{eqnarray}
The integral again defines a gamma function, and so
\begin{eqnarray}
\nu > 0 \, : \qquad\qquad\qquad &&
\nonumber\\
K_\nu(z) &\simeq& { 1 \over 2 \nu } \, 
\left({z \over 2} \right)^{-\nu} ( 1 - \nu \gamma )  \,,
\label{big}
\end{eqnarray}
with again the result containing the correct finite constant as well as
the singular pole term.

The result (\ref{small}) for $\nu < 0 $ can be analytically
continued into the region $\nu> 0$. In this region it involves a
higher power of $z$  than that which appears in the other evaluation
(\ref{big}), and hence this analytic continuation of the leading result
for $\nu < 0 $ into the region $ \nu > 0$ becomes 
sub-leading here.  Similarly, the result (\ref{big}) for $\nu
>0$ may be analytically continued into the region  $\nu < 0$ where it
now becomes sub-leading.  An examination of the defining integral
representation (\ref{int}) shows that these sub-leading analytic
continuation terms are, in fact, the dominant, first-subleading
terms.%
\footnote{For example, subtracting the leading term (\ref{small}) 
for $\nu < 0$ from the integral representation (\ref{int}) gives
$$
K_\nu(z) - {1 \over 2} \left( {z \over 2} \right)^\nu \Gamma( -\nu)
	=
{1 \over 2} \int_0^\infty {dk \over k} k^\nu 
	\left[ e^{-z k /2 } - 1 \right] \, e^{- z / (2k) } \,.
$$
For $ 0 > \nu > -1$, the integral on the right-hand-side of the equation 
converges when the the final exponential factor in the integrand is
replaced by unity, the $z \to 0$ limit of this factor. Hence this 
final factor may be
omitted in the evaluation of the first sub-leading term. A partial
integration presents the result as
$$
{ z \over 4 \nu }  \, \int_0^\infty dk \, k^\nu \, e^{-z k/2} \,,
$$
whose evaluation gives precisely the analytic continuation of the
leading term (\ref{big}) for $\nu > 0$.}
For $\nu > 0$ one term is leading and the other subleading, while 
for $\nu < 0$ their roles are interchanged. Thus their sum
\begin{equation}
K_\nu(z) \simeq { 1 \over 2 \nu} \left\{ 
\left( { z \over 2} \right)^{-\nu} \left[ 1 - \nu \gamma \right] -
 \left( { z \over 2} \right)^{\nu} \left[ 1 + \nu \gamma \right] 
 \right\}
\label{about}
\end{equation}
contains both the leading and the first subleading terms for both 
$\nu > 0$ and $\nu < 0 $. In the limit $ \nu \to 0$ the (``infrared'' 
and ``ultraviolet'') pole terms in this sum
cancel, with the variation of the residues of the poles producing a
logarithm, yielding the familiar small $z$ result 
\begin{equation}
 K_0(z) = - \ln ( z/2) - \gamma \,.
\end{equation}
It must be emphasized that the correct constant terms [$ \ln
2 - \gamma $] are obtained by this method in addition to the logarithm
$ -\ln z$ which is large for small $z$.  The result (\ref{about}) was
derived from the analytic continuation of results that were easy to
compute in one or the another region where either ``infrared'' or
``ultraviolet'' terms dominated. This is the essence of our method.
Of course, the general result (\ref{about}) could be obtained by a
more careful computation of both the leading and first-subleading
terms in either of the separate $\nu > 0$ or $ \nu < 0$ regions as was
done in the previous footnote. In some of the physical
examples that we shall work out, such an extraction of the subdominant
terms would be very difficult indeed, although possible in
principle. Thus our method acquires real power for the physical
problems.

In the physical examples described below, one could object that we do
not explicitly prove that larger subleading terms are not present.  In
the limit of the typical small parameter $z \to 0 $ that concerns us
(with now $\nu \to 3$), for our physical examples we shall
find (after extracting some overall factor of $z$) leading terms of
order $z^{(3-\nu)}$ for $\nu>3$ and $z^{(\nu-3)}$ for $\nu<3$. One
may then ask if an additional term that has a power dependence between
$z^{(\nu-3)}$ and $z^{(3-\nu)}$ can appear.  However, simple
dimensional analysis shows 
that such terms of intermediate order cannot appear. The point is that
these physical examples involve only two different physical mechanisms
that dominate at large and small scales, and that the two different
mechanisms involve different combinations of the basic physical
parameters and hence give quite different dependencies on the small
parameter when the dimension $\nu$ departs from $\nu =
3$. Incidentally, it should go without saying that the physical
examples that we shall study have a basic theoretical description that
is valid for a range of spatial dimensions $\nu$ about $\nu = 3$.

\section{Energy Loss in a Plasma}

The usual method for obtaining the energy loss for a charged particle
moving through matter is to divide the calculation into two parts: The
long-distance, soft collisions and the short-distance, hard
collisions. Collective effects are important in the long-distance
part, and it is evaluated from the ${\bf j} \cdot {\bf E}$ power loss
of a particle moving in a dielectric medium.  The hard collisions are
described by Coulomb scattering. The rub is to join the disparate
pieces together. For the case of classical scattering, this is often
done by computing the energy loss in Coulomb scattering out to some
impact parameter, and then adding the ${\bf j} \cdot {\bf E}$ energy
loss for all larger impact parameters.  Although such methods do yield
the correct large logarithm without much difficulty, the logarithm of
the ratio of the two scales which is large, the purely numerical
constants (which one expects to be of order one) that accompany the
logarithm are harder to compute.  Here we describe an easily applied
method that yields a unique result -- the result including the
constants in addition to the large logarithm.  The new idea is to
compute the energy loss from Coulomb scattering over all angles, but
for dimensions $\nu > 3$ where there are no infrared divergences. A
separate calculation of the energy loss using the ${\bf j} \cdot {\bf
E}$ heating is done for $\nu < 3$, where the volume integration may be
extended down to the particle's position without encountering an
ultraviolet divergence. Both of these results have a simple pole at
$\nu = 3$, but they both may be analytically continued beyond their
initial range of validity.  In their original domain of dimension
$\nu$, both calculations are performed to the leading order in the
plasma density.  As will be seen, although the Coulomb scattering
result is the leading order contribution for $\nu > 3$, it is of
subleading order when $\nu < 3$. Conversely, the ${\bf j} \cdot {\bf
E}$ heating is subleading for $\nu > 3$ but leading for $\nu <
3$. Hence, the sum of the two (analytically continued) processes gives
the leading and (first) subleading terms in the plasma density for all
dimensions $\nu$, and thus, in the limit of this sum at $\nu = 3$, the
pole terms must cancel with the remainder yielding the correct physical
limit to leading order in the plasma density.

The fully ionized, classical plasma with which we are concerned is
described exactly by a coupled set of kinetic equations, the
well-known BBGKY hierarchy as described, for example, in Section 3.5
of ref.~\cite{huang}. We are interested, however, in the computation
to leading order in the plasma density of the energy loss of a fast
particle traversing the plasma.  The correct equations that
govern the leading order low-density behavior change as the spatial
dimensionality $\nu$ changes. For $\nu < 3$, the long-distance,
collective effects dominate, and the formula derived by Lenard and
Balescu applies \cite{len}, \cite{dup}. This formula describes the
interaction of the various species that the plasma may contain.  In
the limit in which one species is very dilute, as is our case in which
we examine the motion of a single, fast  ``test particle''
moving through the plasma, the energy lost in the particle motion is
described by its ${\bf j} \cdot {\bf E}$ Joule heating with the
background plasma response given by the permittivity of a
collisionless plasma.  On the other hand, when the spatial dimension
$\nu$ is greater than $3$, the short-distance, hard Coulomb collisions
dominate. For these dimensions, the leading low density limit of the
BBGKY hierarchy is described by the familiar Boltzmann equation. The
Boltzmann equation is derived, for example, also in Section 3.5 of
ref.~\cite{huang}.  We use the Boltzmann equation to obtain the
leading order energy loss rate when $\nu > 3$.  Again, since we are
concerned with the motion of a single, fast ``test particle'', the
Boltzmann equation reduces to the product of the energy loss weighted
cross section times the plasma density. The derivations that we have
just described, which start from first principles, justify the methods
outlined in the previous paragraph, the methods that we shall use.

Since we are only interested in describing the new method, we simplify
the discussion by treating only the electrons in a classical plasma
(electron recoil gives the dominant energy loss since they are light),
and by taking the moving projectile velocity $v_p$ to be much larger than
the electron velocities in the plasma so that the latter may be
neglected relative to $v_p$. We shall assume, however, that the
projectile velocity is small in comparison with the velocity of light
so that this particle produces a simple Coulomb field (as modified by
the plasma) and that non-relativistic mechanics applies.

\subsection{$\nu < 3$}

We first compute the ${\bf j} \cdot {\bf E} $ heating with $ \nu < 3$.
The electric field ${\bf E}$ is produced by the point projectile of 
charge $e_p$ moving with velocity ${\bf v}_p$ which gives the charge
density
\begin{eqnarray}
\rho({\bf r},t) &=& e_p \, \delta^{(\nu)}({\bf r} - {\bf v}_p t)
\nonumber\\
&=& e_p \, \int { (d^\nu k) \over (2\pi)^\nu} 
	\exp\left\{ i {\bf k} \cdot \left( {\bf r}
	- {\bf v}_p \, t \right) \right\} \,.
\end{eqnarray}
In our non-relativistic limit, the electric field is curl free while 
$
\nabla \cdot {\bf D} = \rho
$,
and so
\begin{equation}
{\bf E}({\bf r},t) =  e_p \, \int { (d^\nu k) \over (2\pi)^\nu} 
{ - i {\bf k} \over k^2 \, \epsilon({\bf k}, {\bf k} \cdot {\bf v}_p ) } 
\, \exp\left\{ i( {\bf k} \cdot {\bf r}
- {\bf k} \cdot {\bf v}_p \, t ) \right\} \,,
\end{equation}
where $ \epsilon({\bf k},\omega)$ is the wavenumber and frequency
dependent electric permittivity of the plasma. 
Note that we use rationalized Gaussian units so that electrostatic
potential of a point charge in three dimensions has the form 
$\phi = e / (4\pi \, r)$.
Since the current of the projectile is given by 
$
{\bf j}({\bf r},t) = {\bf v}_p \, \rho({\bf r},t)
$,
this energy loss mechanism gives 
$ 
d E / dt = - e_p {\bf v}_p \cdot {\bf  E}({\bf v}_p t , t) 
$, 
or
\begin{equation}
{d E_< \over dt } =  e_p^2 \int { (d^\nu k) \over (2\pi)^\nu } 
{ i \over k^2 } 
{ {\bf k} \cdot {\bf v}_p 
\over \epsilon({\bf k}, {\bf k} \cdot {\bf v}_p ) } \,.
\label{jjoule}
\end{equation}
The electric permittivity is the boundary value of an analytic
function, $\epsilon({\bf k}, \omega) = 
\epsilon({\bf k}, \omega + i \eta) $, $ \eta \to 0^+ $, with 
$ \epsilon({\bf k}, z)  =  \epsilon({\bf k}, - z) $ an even function
of $ z $ which also depends only on the modulus $|{\bf k}|$ of 
${\bf k}$ (by rotational invariance). 
[This is demonstrated in the explicit form (\ref{fullfn})
below.] Thus, in view of the factor ${\bf k} \cdot {\bf v}_p$ which
extracts the odd part of the rest of the integrand in
Eq.~(\ref{jjoule}), we may write this energy loss as 
\begin{equation}
{d E_< \over dt } =  e_p^2 \int { (d^\nu k) \over (2\pi)^\nu } 
{ 1 \over k^2 } \, 
{\rm Im} \left\{ { - {\bf k} \cdot {\bf v}_p  \over \epsilon({\bf k}, 
{\bf k} \cdot {\bf v}_p ) } \right\} \,,
\label{joule}
\end{equation}
in which ${\rm Im}$ denotes the imaginary part.

For our calculation to leading order in the plasma density, the 
permittivity function may be taken in the first (one-loop) 
approximation \cite{dirk}
(the ring graph of quantum statistical mechanics)
\begin{equation}
\epsilon({\bf k},\omega ) = 1 - { e^2 \over k^2 } \int {(d^\nu
{\bf p}) \over (2\pi\hbar)^\nu}  \, n_e( {\bf p} ) \,
 { 2 \left[ ( {\bf p} + \hbar {\bf k} )^2 / (2m_e)
	-  {\bf p}^2 / (2m_e) \right]
\over \hbar^2 ( \omega + i \eta)^2  - 
 \left[  ( {\bf p} + \hbar {\bf k} )^2 / (2m_e) 
	-  {\bf p}^2 / (2m_e)  \right]^2 } \,,
\label{fullfn}
\end{equation}
where the $\eta \to 0^+$ in the denominator corresponds to a
retarded response, and where  
\begin{equation}
\int {(d^\nu {\bf p}) \over (2\pi\hbar)^\nu}  \, n_e( {\bf p} ) 
	= n_e 
\end{equation}
is the electron number density, and $m_e$ is the mass of the electron.
Since we are examining the limit in
which the projectile is moving very rapidly with respect to the
average thermal velocity of the electrons, the electron motion in the
plasma may neglected. This corresponds to setting ${\bf p} =0$ in the
remainder of the integrand in Eq.~(\ref{fullfn}), which gives
\begin{equation}
\epsilon({\bf k},\omega) = 1 - { \omega_e^2 \over 
(\omega + i \eta)^2 - ( \hbar k^2 /2m_e)^2 } \,,
\label{recoil}
\end{equation} 
where $\omega_e$ is the plasma frequency for the electrons defined by
\begin{equation}
\omega_e^2 = { e^2 \, n_e \over m_e } \,.
\label{plfq}
\end{equation}
With $\omega = {\bf k} \cdot {\bf v}_p$, two length scales appear, 
$ v_p / \omega_e$ and $\hbar / ( m_e v_p)$. It is easy to check that the
scale for the wave number integration in the energy loss (\ref{joule})
(with, of course, $\nu < 3$) 
is set by the former, classical length. Hence the latter quantum
length appears as a correction involving the dimensionless parameter
$\hbar^2 \omega_e^2 / (m_e v_p^2)^2 $. Since $\omega^2_e$ is
proportional to the electron density, and we are working to leading
order in this density, we must omit this small parameter and use the
purely classical limit%
\footnote{Note that our dimensional regularization method yields the 
correct leading-order result, unaccompanied by any higher-order 
terms that only give a part of the higher-order corrections and thus
represent spurious corrections.} 
\begin{equation}
\epsilon({\bf 0},\omega) = 1 - { \omega_e^2 \over (\omega +
  i\eta)^2 } \,,
\end{equation}

This is the limit to be used in the energy loss 
(\ref{joule}). In this limit,
\begin{equation}
{\rm Im} \left\{ { - \omega \over \epsilon({\bf 0}, \omega) } 
\right\} = \pi \, \omega_e^3 \,
   \delta\left( \omega^2 - \omega_e^2 \right) \,.
\end{equation}
Hence, performing the integration over the component of
${\bf k}$ parallel to ${\bf v}_p$, and writing $dx = v_p dt$
gives
\begin{equation}
{ d E_< \over dx} = {e_p^2 \over 2} \,  \int { (d^{\nu-1} k) \over
  (2\pi)^{\nu-1} } { \omega_e^2 \over \omega_e^2 + v_p^2 k^2 } \,.
\label{wnint}
\end{equation}
Exponentiating the denominator via 
\begin{equation}
D^{-1} = \int_0^\infty ds e^{-sD} \,,
\label{exp}
\end{equation}
interchanging integrals, performing the resulting $\nu - 1$ Gaussian
$k$ integrals, and recognizing the final $s$ integral as a
standard representation of the $\Gamma$ function gives
\begin{equation}
{ d E_< \over dx } = { e_p^2 \over 2}  \, 
\left({ \omega_e^2 \over 4 \pi v_p^2 }
\right)^{\nu -1 \over 2} \,  \Gamma\left( { 3 - \nu \over 2 } \right)
\,,
\label{first}
\end{equation}
or, with the neglect of terms which vanish when $\nu \to 3$,
 \begin{equation}
{ d E_< \over dx } =  { e_p^2 \omega_e^2 \over 4\pi \, v_p^2 } \, 
\left({ \omega_e^2 \over 4 \pi v_p^2 }
\right)^{\nu - 3 \over 2} \,  \left\{ { 1 \over 3 - \nu } - {\gamma
    \over 2} \right\} \,.
\label{below}
\end{equation}
The pole in this expression, which becomes negative when $ \nu > 3$,
corresponds to the ultraviolet divergence which appears when $\nu \to
3$ in the wavenumber integral (\ref{wnint}).

\subsection{$\nu > 3$}

We turn now to the $\nu > 3$ case where the energy loss is computed by
single-particle scattering.  By the conservation of energy, the energy
loss in the scattering of the projectile velocity ${\bf v}_p \to {\bf
v}'_p$ on electrons whose initial velocity may be neglected is 
\begin{equation}
\Delta E =  - {m_p \over 2} \, \left[ {v_p'}^2 - v_p^2 \right] = 
{ m_e \over 2 } \, {v_e'}^2 \,,
\end{equation}
where $v'_e$ is the speed of the scattered
electron.  Since the initial electron has negligible momentum, 
this can be written in the invariant form $ \Delta E = q^2 / ( 2 m_e)$, 
where ${\bf q}$ is the electron momentum transfer in the scattering 
process. With the initial electron at rest, the differential rate of 
scattering is $ v_p n_e d\sigma $, where $n_e$ is the electron density 
in the plasma and $d\sigma$ is the cross section element. Since 
$dx = v_p dt$, the energy loss for $\nu > 3$ is given by 
\begin{equation}
{d E_> \over dx} = { n_e \over 2 m_e} \, 
\int d\sigma \, q^2 \,.
\label{scatt}
\end{equation}

We first evaluate this scattering contribution when the interaction is
weak, when $\eta = e_p e / \hbar v_p \ll 1$. In this case, the
quantum-mechanical Born approximation result is appropriate with, in
$\nu > 3$ dimensions,
\begin{equation}
\int d\sigma_B \, q^2 = \int{ (d^\nu {\bf p}') \over (2\pi\hbar)^\nu }
\, 2 \pi \hbar \, 
\delta \left( { {p'}^2 \over 2m} - { p^2 \over 2m } \right) 
\left( { \hbar e_p e \over q^2 } \right)^2 {1 \over v} \, q^2
\,.
\label{cross}
\end{equation}
Here $ ( 1 / m) = ( 1/ m_e ) + ( 1 / m_p ) $ defines the reduced 
mass $m$ and $v$ is the relative velocity between the electron and the
projectile. Writing
$
q^2 = 4 \, m^2 v^2 \sin^2\theta/2  \,,
$
and
\begin{equation}
(d^\nu {\bf p}') = m \, {p'}^{(\nu-2)} \,  d ( {p'}^2 /2m ) 
\, \Omega_{\nu-2} \, \sin^{\nu -2} \theta \, d\theta \,,
\end{equation}
with 
$
\sin^{\nu-2} \theta = [2 \cos\theta/2 \, \sin\theta/2 ]^{\nu -2} \,,
$
and noting that the solid angle $\Omega_{\nu-2}$ is given by 
\begin{equation}
{ \Omega_{\nu-2} \over 2\pi} = 
{ \pi^{(\nu-3)/2} \over \Gamma \left( {\nu -1 \over 2} \right) } \,,
\end{equation}
we get, on setting $\chi = \theta/2$,
\begin{equation}
\int d\sigma_B \, q^2 = { (e_p e)^2 \over 2\pi \, v^2 } \left( { m^2 v^2
    \over  \pi \hbar^2 } \right)^{(\nu -3)/2} { 1 \over \Gamma \left(
    { \nu -1 \over 2} \right) } \int_0^{\pi/2} d\chi \, \cos^{\nu -2}\chi
       \, \sin^{\nu-4}\chi  \,.
\end{equation}
The integral which appears here has the value $ (\nu - 3)^{-1} + O(\nu
- 3) $ as one can show by dividing it into two parts with a suitable
partial integration or by expressing it in terms of the standard
integral representation of the Beta function.  Since we neglect the
motion of the initial electron, the relative velocity $v$ may be
replaced by the projectile velocity $v_p$, and so using the result in
Eq.~(\ref{scatt}) gives
\begin{equation}
{d E_>^{(Qm)} \over dx} = { e_p^2 \omega_e^2 \over 4\pi \, v_p^2 } 
\left( { m^2 v_p^2
    \over \pi \hbar^2 } \right)^{\nu -3 \over 2} \left\{ { 1 \over \nu -3 } +
  { \gamma \over 2} \right\} \,.
\label{qabove}
\end{equation}
The pole in this expression, which becomes negative when $ \nu < 3$, 
corresponds to the infrared divergence of the momentum integral
(\ref{cross}) in the $\nu \to 3$ limit.

\subsection{$\nu = 3$}

When the result (\ref{qabove}) is added to that in Eq.~(\ref{below}) 
the divergent
pole terms cancel, and the physical limit $ \nu \to 3$ is
\begin{equation}
{d E_{Qm} \over dx} = { e_p^2 \omega_e^2 \over 4\pi \, v_p^2 } \ln
\left( { 2m v_p^2 \over  \hbar \omega_e  } \right) \,.
\label{qlimit}
\end{equation}
For small $\eta$, this is the correct result to leading order in
the plasma density.  Instead of using the plasma density for the
demonstration, it is equivalent to use the linearly related plasma 
frequency $\omega_e^2$.  We have computed the leading and subleading 
terms in this quantity. The result (\ref{below}) for $d E_< / dx$ 
involves $\omega_e^2 \times \omega_e^{(\nu-3)} $ while the result
(\ref{qabove}) for $ d E_> /dx$ involves just $\omega_e^2$. Hence, for
$\nu < 3$, (\ref{below}) is leading and (\ref{qabove}) is subleading,
while for $\nu > 3$, their roles are reversed.  Thus, in either region
the sum of the two contributions contains both the leading and (first)
subleading terms, and so the limit of the sum at the physical
dimension $\nu = 3$ yields the correct result to leading order in the
plasma density.

The result (\ref{qlimit}), including the proper constants inside the
logarithm, may also be essentially obtained by applying the ${\bf j} \cdot
{\bf E}$ heating formula (\ref{joule}) directly in three dimensions
with the use of the single-ring graph quantum form (\ref{recoil}) of
the dielectric function in the limit in which the electrons in the
plasma are taken to have negligible velocity. Placing this function in
Eq.~(\ref{joule}) with $\nu = 3$, writing ${\bf k} \cdot {\bf v}_p = 
\cos \theta$ in the resulting delta function, and using this delta
function to eliminate the polar angle $\theta$ of the solid angle
integration gives a remaining integral over the magnitude $k$ of
$|{\bf k}|$, $ \int dk / k$.  The leading terms for small $\omega_e$ 
of the upper and lower limits of this logarithmic integral
give the result (\ref{qlimit}), except that
the correct reduced mass $m$ in Eq.~(\ref{qlimit}) is replaced by the
electron mass $m_e$ since the current ${\bf j}$ describes the motion
of a very heavy projectile particle.  This sort of calculation was
done some time ago by Lindhard \cite{lind}. Although the reduced mass
correction is negligible when the projectile is a heavy ion, it does
represent a conceptual shortcoming of the quantum-corrected, joule
heating treatment.  Moreover, this treatment completely breaks down 
when the projectile is itself an electron.  This sort of
dielectric treatment is also
restricted to the case of a cold plasma whose electron velocities are
much less than that of the projectile. On the other hand, our method
is easily extended \cite{sawyer} to treat the case of a hot plasma
where this restriction is not imposed, and again a complete
calculation can be performed which includes the constants in addition
to the logarithm.

Although, as we have just seen, the ${\bf j} \cdot {\bf E}$
calculation can be improved to obtain the correct energy loss (except
for the replacement of the reduced by the electron mass), with the
computation always done in three dimensions, we do not know of a
similar improvement of the Boltzmann equation in three dimensions
which yields the correct result.  One might be tempted to replace
the Coulomb potential by the screened Debye
potential. This alteration changes the $1 / {(q^2)}^2 $ factor in the
cross section formula (\ref{cross}) by
\begin{equation}
\left( { 1 \over q^2 } \right)^2 \to 
\left( { 1 \over q^2 + \hbar^2 \, \kappa^2 } \right)^2 \,,
\end{equation}
in which $\kappa^2 = e^2 \, n / \, T$ is the squared Debye wave number
for the plasma.  This alteration removes the long-distance infra-red
divergence, and the cross section formula (\ref{cross}) now converges
in three dimensions. Exponentiating this denominator using the
integral representation (\ref{exp}) with an additional factor of $s$
in the integrand to produce the square makes the remainder of the
calculation easy, and one finds that 
\begin{equation}
{d E_{D} \over dx} = { e_p^2 \omega_e^2 \over 4\pi \, v_p^2 } \ln
\left[ \left( { 2m v_p \over  \hbar \kappa  } \right) - {1\over2} 
\right]  \,.
\end{equation}
Although the constant out in front of the logarithm is again the correct
over-all constant, the argument of the logarithm is quite different
from the correct form given in Eq.~(\ref{qlimit}). This should have
been expected at the outset because  Debye screening describes the static
screening of a particle at rest in the plasma, not a dynamical
screening of a fast moving particle which is the case that we are 
examining.  As far as I know, such a dynamical screening within a
Boltzmann equation context cannot be done.

Our method can be used to extend the result (\ref{qlimit}) to arbitrary
values of $\eta = e e_p /(4\pi \, \hbar v_p) $, always retaining the correct
additional constants.  To do this,  we use some clever  mathematics of
Lindhard and  Sorensen \cite{lind2}, but  in a  manner which justifies
that these constants have been kept. Namely, we compute
\begin{equation}
\Delta {d E_> \over dx} = { n_e \over 2 m_e} \, 
\int \left( d\sigma - d\sigma_B \right) \, q^2 \,.
\end{equation}
This difference is well behaved in the limit $\nu \to 3 $ since the
pole at $\nu = 3$ produced by the cross section integral comes from
soft, infrared physics which is completely contained in the Born
approximation $d\sigma_B$. Hence the three-dimensional partial wave
decomposition of the scattering amplitude may be used, and then 
standard manipulations yield
\begin{eqnarray}
\int \left( d\sigma - d\sigma_B \right)\, q^2 
&=& 2 \pi \hbar^2 \sum_{l=0}^\infty (l+1)
\Bigg\{ \left[2 - e^{2i [\delta_l - \delta_{(l+1)}] } -
  e^{- 2i [\delta_l - \delta_{(l+1)}] } \right]
\nonumber\\
&& \qquad\qquad\qquad
- \left[2 - e^{2i [\delta_l - \delta_{(l+1)}] } -
  e^{- 2i [\delta_l - \delta_{(l+1)}] } \right]_B
\Bigg\} \,.
\end{eqnarray}
For the Coulomb potential
\begin{equation}
e^{2i\delta_l} = { \Gamma(l+1+i\eta) \over  \Gamma(l+1-i\eta) } 
e^{i\phi} \,,
\end{equation}
where the phase $\phi$ is independent of $l$. Using $\Gamma(z+1) = z
\Gamma(z)$ and a little algebra, we find that
\begin{eqnarray}
\int \left( d\sigma - d\sigma_B \right) q^2 &=&
4 \pi \eta^2 \hbar^2 \sum_{l=0}^\infty 
\left[ { 1 \over l+1+i\eta} + { 1 \over l+1-i\eta } 
- { 2 \over l+1} \right]
\nonumber\\
&=& -  \, {e^2 e_p^2 \over 4\pi \, v_p^2 } \,  2
\left[ {\rm Re} \, \psi(1+i\eta) + \gamma \right] \,,
\end{eqnarray}
where $\psi(z)$ is the logarithmic derivative of the gamma function,
$\psi(z) = \Gamma'(z) / \Gamma(z) $, and Re denotes the real part.
Recalling the definition (\ref{plfq}) of the plasma frequency, we now
have \cite{bloch}
\begin{equation}
\Delta {dE_> \over dx} = - {e_p^2 \omega_e^2 \over 4\pi \, v_p^2 } 
\left[ {\rm Re} \, \psi(1+i\eta) + \gamma \right] \,,
\label{inter}
\end{equation}
with the energy loss for all $\eta$ values given by
\begin{eqnarray}
{dE \over dx} &=&  {dE_{Qm} \over dx} +
        \Delta {dE_> \over dx} 
\nonumber\\
	&=&
  {e_p^2 \omega_e^2 \over 4\pi \, v_p^2 } \left\{ 
	\ln\left( { 2 m v_p^2 \over \hbar \omega_e} \right) 
-\left[ {\rm Re} \, \psi(1+i\eta) + \gamma \right] \right\} \,.
\label{sum}
\end{eqnarray}

In the classical case, $\eta = e e_p / (4\pi \,  \hbar v_p) $ 
becomes large. Using the limit
\begin{equation}
|z| \to \infty \,: \qquad \psi(1+z) = \ln z + O(z^{-1}) \,,
\end{equation}
Eq.~(\ref{sum}) yields the classical form
\begin{equation}
{ d E_{Cl} \over dx} = { e_p^2 \omega_e^2 \over 4\pi \, v_p^2 } 
\ln\left(2 e^{-\gamma} {4\pi \,   m v_p^3 \over e_p e \omega_e }\right)
    \,.
\label{climit}
\end{equation}
This result, including the proper constant $ 2 e^{-\gamma}$ that
appears within the logarithm, was obtained long ago by Kramers
\cite{kramers}.  It may also be obtained directly \cite{brown3} with
our dimensional continuation methods by using the classical Coulomb
scattering cross section for dimension $\nu > 3$ in the scattering
energy loss expression (\ref{scatt}).

\section{Lamb Shift}

Essentially the same method applied here has been used before in my
Quantum Field Theory book \cite{brown} to calculate the Lamb shift for
hydrogen-like atoms, with the small parameter role of the plasma
density replaced by the nuclear charge $Ze$. That exposition, however,
was presented in a somewhat mystical manner, and it unfortunately did
not bring out the essence of the method. This will be rectified now
and the process will provide another example of how the method works.
Section 8.7 of ref.~\cite{brown} explains in detail how a radiative
energy correction may be expressed as a matrix element of the electron
self-energy operator\footnote{In this formulation, the vacuum
polarization contribution to the Lamb shift appears as a separate
modification of the Coulomb potential.  This is a simple correction
which does not involve an interplay between long and short distances
that concerns us in this paper, so we omit the effect of vacuum
polarization here.}  $\Sigma(E)$ in Coulomb, bound-state Dirac wave
functions of energy $E$.  The Lamb shift is an energy difference that
has both infrared and ultraviolet contributions just as in the more
elementary plasma energy loss example explained above.  In $\nu$
spatial dimensions, these become two distinct physical processes that
scale in different ways with a characteristic atomic
energy\footnote{Simple dimensional analysis shows that characteristic
atomic energy in $\nu$ spatial dimensions is given by $ {\cal E} =
(\hbar^2 / m) \, (Z^2 e^4 \, m^2 / \hbar^4 )^{(4-\nu)^{-1}} $ , which
reduces to the familiar scale $ {\cal E} = Z^2 e^4 \, m / \hbar^2 $ in
three dimensions. Since the scaling behavior of $\Sigma(E)$ is more
simply expressed in terms of ${\cal E}$ rather than a dimensionless
parameter formed from $Ze^2$, we use ${\cal E}$ as our small
parameter.} ${\cal E}$: Removing a common overall factor, the
ultraviolet contribution behaves as $ {\cal E}$, while the infrared
contribution goes as ${\cal E}^{\nu - 2}$.  Since the energy ${\cal
E}$ vanishes when $Ze^2$ vanishes, we may take ${\cal E}$ (implicitly
divided by some fixed energy scale to yield a dimensionless number) as
our small parameter. Thus, just as in the previous plasma case, the
infrared part dominates when $\nu < 3$, the ultraviolet part dominates
when $\nu > 3$, the sum of the two contributions analytically extended
in the vicinity of $\nu = 3$ always contains both the dominant and
leading sub-dominant terms, and so the $\nu \to 3$ limit of this sum
yields the correct leading-order Lamb shift.  To tame the infrared
divergences which are prevalent when $\nu < 3$, the binding of the
electron must be accounted for.  To tame the ultraviolet divergences
that may appear when $\nu>3$, a relativistic treatment must be made.
We turn now to sketch this calculation. The detailed expressions that
we shall need are derived and presented in ref.~\cite{brown}.

\subsection{ $ \nu < 3 $ }

\begin{figure}[t]
    \begin {center}
    \leavevmode
    
    \epsfbox {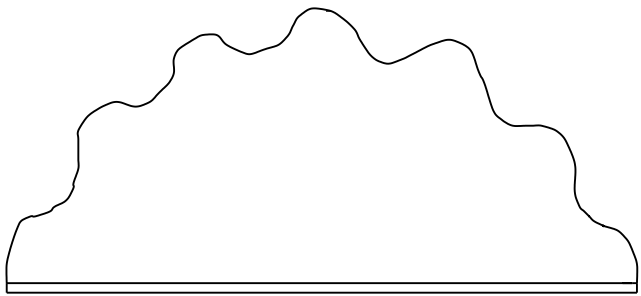}
    \end {center}
\caption
    {%
	Feynman diagram for the electron self-energy operator of
	Eq.~(\ref{intless}). The waving line stands for the transverse
	virtual photon exchange of the radiation gauge. The narrowly
	spaced straight lines stand for the non-relativistic electron 
	(operator) propagator in the nuclear Coulomb field.
    }
\end {figure}

With soft (virtual) photon exchange, the leading terms are given by
the non-relativistic limit of the electron motion. In this
non-relativistic limit, the calculation is most easily performed in
the radiation gauge. The Coulomb self-energy contribution is removed
by a mass renormalization. The photon exchange contribution to the
electron self-energy is properly renormalized by a subtraction so that
it vanishes for a free particle. An elementary computation yields
\begin{equation}
\Sigma_<(E) = e^2 \, \int 
{(d^\nu k) \over (2\pi)^\nu} \, 
\left( \delta_{lm} - \hat k_l \hat k_m \right) { 1 \over 2 k^2} \,
{ {\bf p} \over m c} \cdot 
{ H - E \over H - E + k\, \hbar c - i \epsilon } \, 
{ {\bf p} \over m c} \,.
\label{intless}
\end{equation}
Here $H$ is the non-relativistic Hamiltonian for the hydrogen-like 
atom with nuclear charge $Ze$. 
This result, which may be obtained from old-fashioned second-order 
time-dependent perturbation theory, involves the atomic Coulomb 
exchange to all orders as shown in Fig.~1.
It is just Eq.~(5) of the original Lamb shift 
paper of Bethe \cite{bethe} except that it is written in $\nu$ rather 
than $3$ dimensions.  It is also essentially Eq.~(8.7.43) of  
ref.~\cite{brown}. Performing
the integrations (as are explicitly done in the ref.~\cite{brown}) 
gives, with the neglect of terms that vanish at $\nu = 3$, 
\begin{equation}
\Sigma_<(E) = {2 \over 3\pi } \, { e^2 \over 4 \pi \, \hbar c} 
\left[ { 1 \over 3 - \nu }
 + { 5 \over 6} - {\gamma \over 2} \right] \,
{ {\bf p} \over m c} \cdot 
\left( H - E \right) \left[
{ H - E  - i \epsilon \over  \sqrt{\pi} \, \hbar c } 
\right]^{\nu -3} \,  { {\bf p} \over m c} \,.
\end{equation}
In describing the scaling of the results with respect to the small
parameter ${\cal E}$, we implicitly consider matrix
elements of the self-energy operator in a bound-state energy 
eigenfunction and omit the scale associated with the two ${\bf p}$ 
operators that always appear in the expressions. Here, the 
two ${\bf p}$ flank the operator $(H-E)^{\nu - 2}$ which has the
characteristic atomic size ${\cal E}^{\nu -2}$. The operator 
$(H-E)^{\nu - 2}$ has 
this typical scale for any intermediate state when a complete set of 
intermediate states are inserted within the matrix element. Since the
whole expression converges, it has the size ${\cal E}^{\nu -2}$.  
Thus we confirm the that the leading term for $\nu < 3$ 
in the Lamb shift is of order ${\cal E}^{\nu -2 }$ as stated before.
The divergence that appears when 
$\nu$ approaches $3$ is, in view of the structure of the integral 
(\ref{intless}), an ultraviolet divergence.

\subsection{ $ \nu > 3 $ } 

\begin {figure}[t]
    \begin {center}
    \leavevmode
    
    \epsfbox {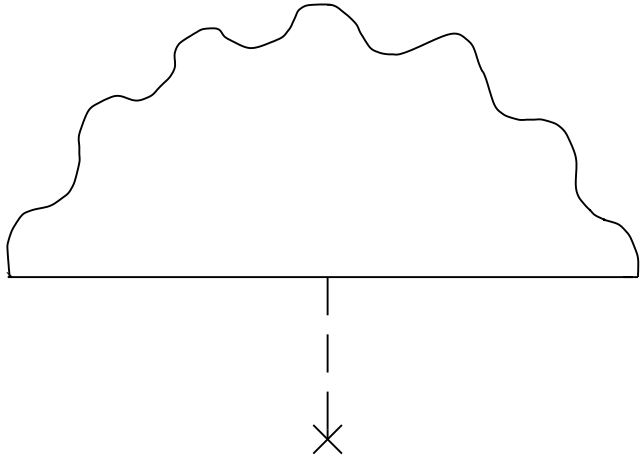}
    \end {center}
\caption
    {%
	Feynman diagram for the electron self-energy operator
	corrected by a single Coulomb exchange with the
	nucleus. The waving line stands for the transverse
	virtual photon exchange of the radiation gauge. The dashed
	line ending in a cross stands for the single Coulomb exchange 
	with the nucleus. 
    }
\end {figure}

In leading order, for spatial dimensionality greater than three, the
leading correction to the electron self energy involves a single Coulomb
interaction between the electron and the nucleus as shown in Fig.~2.
The correction
entails the radiative modification of this interaction which are
described by the order $q^2$ term in the electric form factor $F_1(q^2)$
of the electron, $F_1^\prime(0)$ (the charge radius), and a 
relativistic anomalous magnetic moment effect 
given by the magnetic form factor at zero momentum transfer, $F_2(0)$.
The $q^2$ wave number factor gives the Laplacian of the Coulomb
potential, $ \nabla^2 \, V$. Using the Dirac equation to reduce the
$F_2(0)$ contribution yields a term involving $ \nabla^2 V$ and a
spin-orbit interaction. The Schr\"odinger energy eigenstate matrix 
elements of the operator $\hbar^2 \, \nabla^2 \, V$ are the same as 
those of the
operator $ 2 \, {\bf p} \cdot ( H - E ) \, {\bf p} $. The results are 
derived in ref.~\cite{brown}
and presented there in Eq's.~(8.7.31) and (8.7.36), which we now write
together as 
\begin{eqnarray}
\Sigma_>(E) &=& {2 \over 3\pi } \, {e^2 \over 4 \pi \, \hbar c} \,
\left[ { 1 \over \nu - 3 }
 + {\gamma \over 2} \right]  
\left[ { m c \over   2 \,   \sqrt \pi \, \hbar } \right]^{\nu - 3} 
\, { {\bf p} \over m c } \cdot 
\left( H - E \right)  { {\bf p} \over m c}
\nonumber\\ 
&& \qquad\qquad\qquad\qquad
+ {e^2 \over 4 \pi \, \hbar c } \,
{ 1 \over 4 \pi \, m^2 c^2 } 
\left[ { m c\over  2 \, \sqrt \pi \, \hbar } \right]^{\nu - 3} 
\, \sigma \cdot i[{\bf p}, V ] \times {\bf p} \,,
\label{tired}
\end{eqnarray}
where again terms that vanish at $\nu=3$ are omitted. 
Note that, with our conventions, this result scales as ${\cal E}$,
again confirming an assertion made before.  The
divergence in the first term on the right-hand side of
Eq.~(\ref{tired}) when $\nu$ approaches $3$ comes from 
the contribution of $F_1^\prime(0)$ which contains an infrared
divergence in three spatial dimensions \cite{brown}.

\subsection{ $ \nu = 3 $ }

Since the two effects that we have listed contain the leading and
first sub-leading terms for $\nu$ near $3$, their sum
\begin{equation}
\Sigma(E) = \Sigma_<(E) + \Sigma_>(E)
\end{equation}
evaluated at
$\nu = 3$ must produce the Lamb shift to leading order. Indeed, the
pole terms cancel as they must with the variation of the residues
producing a logarithmic contribution, and one finds that
\begin{eqnarray}
\Sigma(E) = {2 \alpha \over 3\pi } \, 
\ { {\bf p} \over mc} \cdot \left( H - E \right)  
\left\{ \ln \left[ { m c^2 \over  2\left( H - E - i \epsilon \right) }
\right] + { 5 \over 6 } \right\}
{ {\bf p} \over mc}
+ 
{ \alpha \over 4 \pi \, m^2 c^2 } 
\, \sigma \cdot \hbar\nabla V \times {\bf p} \,,
\end{eqnarray}
where now $\alpha = e^2 / (4 \pi \, \hbar c) \simeq 1 / 137 $ may now
be identified with the fine structure constant. 
This is the familiar form\footnote{The imaginary part gives the width
or lifetime of the level.} of the Lamb shift operator. This form
appears in Eq.~(8.7.63) in Ref.~\cite{brown}, and its consequences are
explained there. The
correct factor of $ 5/6 $ has an interesting history
in the computation of the Lamb shift, as related in footnote 13 of
ref.~\cite{feyn}.

It is worth noting that this calculation of the Lamb shift using our
dimensional continuation method is simpler than that done using the
methods of effective quantum field theory \cite{cas} which would
entail an additional matching calculation. It is much simpler than the
conventional, old-fashioned method which utilizes a fictitious photon
mass and a cumbersome joining process with an intermediate,
non-covariant photon momentum cutoff.\footnote{This hoary procedure is
still presented in detail in modern texts on quantum field
theory. See, for example, Section 7-3-2 of Itzykson and Zuber
\cite{itz}, or Section 14.3 of Weinberg \cite{wein}.}

\bigskip\bigskip

\section*{Acknowledgments}

This presentation of my ideas has been improved by conversations with
L. G. Yaffe.  G. Bertsch brought the work \cite{lind} of Lindhard to
my attention and showed me an alternative derivation of his
result. The manuscript was improved by heeding comments of G. Moore. 
This work was supported, in part, by the U. S.  Department of Energy
under grant DE-FG03-96ER40956, and it was largely completed at the Santa
Barbara Institute for Theoretical Physics and at the Los Alamos
National Laboratory.

\end {document}